\documentclass[apj]{emulateapj}
\usepackage{amsmath}
\newcommand{\rsub}{R_\mathrm{sub}}

\shorttitle{Dust Reverberation Mapping and Cosmology}
\shortauthors{S.~F. H\"onig}

\journalinfo{accepted by ApJ Letters (February 17, 2014)}
\submitted{--- arXiv preprint version (\today) ---}

\begin{document}

\title{Dust reverberation mapping in the era of big optical surveys \\
  and its cosmological application}

\author{Sebastian F. H\"onig}
\affiliation{Dark Cosmology Centre, Niels-Bohr-Institute, University of Copenhagen, Juliane-Maries-Vej 30, 2100 Copenhagen \O, Denmark; shoenig@dark-cosmology.dk}

\begin{abstract}
The time lag between optical and near-infrared (IR) flux variability
can be taken as a means to determine the sublimation radius of the
dusty ``torus'' around supermassive black holes in active galactic
nuclei (AGN). I will show that data from big
\textit{optical} survey telescopes, e.g. the \textit{Large Synoptic
  Survey Telescope (LSST)}, can be used to measure dust sublimation
radii as well. The method makes use of the fact that the Wien tail of
the hot dust emission reaches into the optical and can be reliably
recovered with high-quality photometry. Simulations show that dust
sublimation radii for a large sample of AGN can be reliably
established out to redshift $z\sim 0.1-0.2$ with the
LSST. Owing to the ubiquitous presence of AGN up to high redshifts,
they have been studies as cosmological probes. Here, I discuss how
optically-determined dust time lags fit into the suggestion of using
the dust sublimation radius as a ``standard candle'' and propose and
extension of the dust time lags as ``standard rulers'' in combination
with IR interferometry.
\end{abstract}


\setcounter{footnote}{0}

\section{Introduction}

The near-infrared (IR) light curves of active galactic nuclei (AGN)
show variability that is consistent with the optical light
curves, however lagging by tens to hundreds of days
\citep[e.g.][]{Cla89,Gla92,Okn99a,Gla04,Sug06,Kos09}. The lag supposedly
represents the distance from the central engine to the region where the
temperature drops to about 1500\,K so that dust can marginally
survive. This sublimation radius, $\rsub$, has been found to scale with the
square-root of the AGN luminosity using both time delay
  \citep[e.g.][]{Okn01,Min04,Sug06,Kis11a} and IR interferometric measurements\citep[e.g.][]{Kis11a}, as
expected from dust in local thermal equilibrium \citep[e.g.][]{Bar87}.

Both IR interferometry and dust reverberation mapping are limited by
  their technical requirements of either sensitive long-baseline
  arrays or simultaneous availability of optical \textit{and} IR
  instrumentation. Here, I will show that
upcoming large \textit{optical} surveys can recover time lags from hot dust
emission for a huge number of AGN using optical bands,
therefore overcoming some of these limitations. 

Since AGN are ubiquitous in the universe, they may be
  attractive targets as cosmological probes. Both the
broad-line region (BLR) radius \citep{Haa11,Wat11,Cze13} and hot-dust radius
\citep[e.g.][]{Kob98,Okn99b,Okn02,Yos04} seem promising ``standard
candles'', based on the observationally well-established
BLR lag-luminosity \citep[e.g.][]{Kas00,Pet04} and NIR lag-luminosity
relations \citep[e.g.][]{Okn01,Min04,Sug06}. For the BLR,
\citet{Elv02} proposed that a combination of interferometric
observations of emission lines and lag times may
be used as ``standard rulers'', thus bypassing the cosmic distance ladder. Here,
I suggest that the hot-dust lags and near-IR interferometric sizes can
also serve as standard rulers, although requiring advancements in IR interferometry.

\section{Principles of dust reverberation mapping at optical
  wavelengths}\label{sec:idea}

Dust around AGN absorbs the UV/optical radiation from the
  putative accretion disk and reemits in the IR. At about 1500\,K, the
  dust sublimates, corresponding to the hottest dust emission peaking at
  $\sim2\,\micron$. Despite the exponential decrease of the Wien tail, some contribution of the
  hot-dust emission will reach into optical wavebands. Indeed, such a
  dust contribution has been reported by \citet{Sak10} in the
  $I$-band from analyzing the color variability of optical
  variability. Therefore, the optical emission consists of
  contributions from two different emission regions, leading to a
  relative lag.

Dust is considered to be in local thermal equilibrium (LTE), allowing us to
calculate the dust temperature/emission directly
from the absorbed incident radiation. In
\citet{Hon11b}, we presented a theoretical framework showing that
these LTE considerations can be used to calculate temperature
changes $\mathrm{d}T$ for variable incident radiation $\mathrm{d}L$
as $\mathrm{d}T/T = 1/4\,\mathrm{d}L/L$\,\footnote{Without loss of generality, I will
assume that the dust emission follows a black body \citep[see also][]{Kis07,Hon10b,Kis11b,Mor11}.}. Based
on this relation, changes in the temperature
and emission of a pre-defined dust distribution can be calculated to
derive near-IR light curves \citep{Hon11b}.

The optical emission is dominated by the AGN central engine's ``big blue bump'' (BBB). The BBB spectral energy
can be approximated as $\lambda F_\lambda \propto \lambda^{-4/3}$
\citep[e.g.][]{Elv94,Ric06,Ste12a}. For the purpose of this study, I will assume
that the host galaxy is constant and readily removable by image
decomposition and that broad- and narrow-lines do not contribute
significantly ($\la1-10\%$) to the broad-band fluxes (see Sect.~\ref{sec:real}). Therefore, the AGN
emission in the optical can be approximated by
\begin{align}
F_\mathrm{AGN} &= F_V \cdot
                        \left(\frac{\lambda}{0.55\,\micron}\right)^{-7/3}
                        \\ \nonumber
                     &+ \frac{F_\mathrm{BBB}(1.2\,\micron)}{\pi
                          B_\mathrm{1.2\,\micron}(1400\,K)} \cdot \pi
                        B_\mathrm{\lambda}(1400\,K)
\end{align}
$F_V$ denotes the $V$-band flux at $0.55\,\micron$,
$F_\mathrm{BBB}(1.2\,\micron)$ is the BBB flux component at
1.2\,$\micron$, and $\pi B_\mathrm{1.2\,\micron}(1400\,K)$ represents
the flux of a 1400\,K black body at 1.2\,$\micron$ \citep[observed
near-IR color temperature;][]{Kis07,Kis11b}. The
normalization accounts for the fact that AGN show a generic
turnover from BBB-dominated emission to host-dust emission at
about $1-1.2\,\micron$ \citep[e.g.][]{Neu79,Elv94}. 

\begin{figure}
\begin{center}
\epsscale{1.2}
\plotone{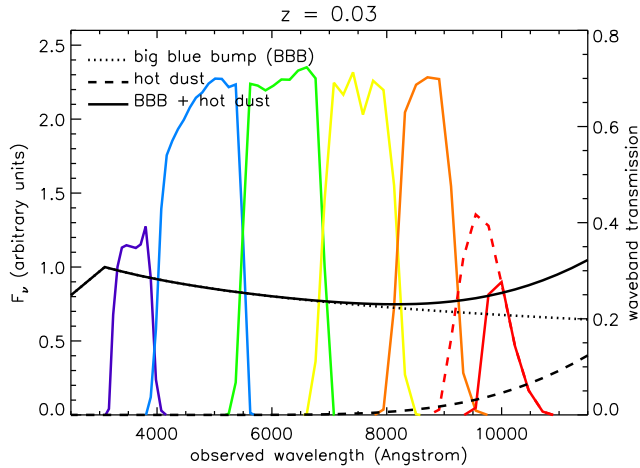}
\caption{AGN template SED for a redshift of $z=0.03$ in the wavelength range from
  $2500-20\,000\AA$. The dotted line is the big-blue bump
  component and the dashed line represents a black-body with temperature
  $T=1500\,$K. The solid line is the combination of both emission
  components. The colored solid lines illustrate the transmission of
  the LSST $u$, $g$, $r$, $i$, $z$, and $y3$ wavebands (violet to red), respectively. The
dashed-red line shows the alternative $y4$ filter.}\label{fig:agn_sed}
\end{center}
\end{figure}
 
\begin{table}[t]
\begin{center}
\caption{Relative contributions of hot dust to wavebands at different redshifts.\label{tab:hotdustcont}}
\begin{tabular}{l c c c c c}
\tableline\tableline
redshift    & $z=0$  & $z=0.05$ & $z=0.1$ & $z=0.2$ & $z=0.3$ \\ \tableline
$i$ band  &  0.019 & 0.012 & 0.007 & 0.003 & $\ldots$ \\
$z$ band &  0.073 & 0.052 & 0.031 & 0.014 & 0.004 \\
$y$ band ($y3$) & 0.206 & 0.158 & 0.109 & 0.053 & 0.020 \\
$y$ band ($y4$) & 0.168 & 0.126 & 0.085 & 0.041 & 0.015 \\\tableline
\end{tabular}
\end{center}
\end{table}

 Fig.~\ref{fig:agn_sed} shows
the total AGN SED, the BBB, and hot-dust
components for a simulated object at redshift $z=0.03$. Overplotted are the transmission curves for the LSST
filters $u$, $g$, $r$, $i$, $z$, and $y$, where the latter may
either be represented by the $y3$ (referred to as $y$ in the following) or $y4$ filter. The
Wien tail of the hot dust reaches into the $z$ and $y$ bands. However, the
fractional contribution of the dust is very
sensitive to the object's redshift. In Table~\ref{tab:hotdustcont},
hot-dust contributions to the total flux in $i$, $z$, and $y$
are listed for $0 < z < 0.3$. Out to about $z\sim0.1$ the dust
contribution to the $y$ band is $\ga$10\% and drops
to $\sim$5\% at $z=0.2$. Therefore, the dust component may be detected
above the BBB out to $z\sim0.1-0.2$.

\section{A dust reverberation mapping experiment for optical surveys\label{sec:sim}}

In this letter I propose to use optical
wavebands for dust reverberation mapping of AGN. Suitable telescope projects are currently
being explored or under construction. The most promising
survey for this experiment will be the
\textit{LSST}. Its cornerstones are high photometric
quality ($<$1\%), high cadence, and multi-year operation. The
feasibility for \textit{LSST} will be illustrated in the following. It can be
easily translated to other surveys.

\subsection{Simulation of observed light curves and construction of a
  mock survey\label{sec:simobs}}

First, it is necessary to simulate survey data and find a method that
allows for recovering dust time lags. \citet{Kel09} show that the optical
variability is well reproduced by a stochastic model based on a
continuous autoregressive process \citep[Ornstein-Uhlenbeck process;
see also][]{Kel13}. The model consists of a
white noise process with characteristic amplitude $\sigma$ that drives exponentially-decaying variability
with time scale $\tau$ around a mean magnitude $m_0$. The
parameters $\sigma$ and $\tau$ have been found to scale with black
hole mass $M_\mathrm{BH}$ and/or
luminosity $L$ of the AGN \citep[e.g.][]{Kel09,Kel13}. For the
simulations, $L$ and $M_\mathrm{BH}$ are chosen and $\sigma$ and $\tau$ are drawn from
the error distribution of the respective relation given in
\citet{Kel09}. Since the amplitude of variability
depends on wavelength, it was
adjusted by $\sigma(\lambda) = \sigma \times
(\lambda/5500\,\mathrm{\AA})^{-0.28}$ as empirically found by \citet{Meu11}.

The BBB light curves are then propagated outward into the
dusty region. Its inner edge, $\rsub$, and thus the dust time lag
$\tau(\rsub)$, scales with $L$ as $\rsub \propto \tau(\rsub) \propto
L^{1/2}$. The reaction of the dust on BBB variability
is modeled using the principles outlined in Sect.~\ref{sec:idea}. For that, the dust is
distributed in a disk with a surface density distribution $\Sigma(r)
\propto r^a$, and the temperature of the dust at distance $r$ from the
AGN is calculated using the black-body approximation for LTE, $T(r) = T_\mathrm{sub}
\times (r/\rsub)^{-0.5}$ (sublimation temperature
$T_\mathrm{sub}=1500\,$K). The
power law index $a$ represents the compactness of the dust
distribution ($a$ very negative = compact; $a\sim0$ = extended) and
results in smearing out the variability signal/transfer
function. Since its actual value is rather unconstrained, a random
value is picked in the range $-2.5 < a < -0.5$, motivated by observations
\citep{Hon10a,Kis11b,Hon12,Hon13}. Using the dust variability model on actual data showed that only a fraction
of the incident variable BBB energy, $w_\mathrm{eff}$, is converted into hot-dust
variability \citep[for details see][]{Hon11b}. Thus, a random $w_\mathrm{eff}$ is picked in the interval
$w_\mathrm{eff} \ \epsilon \ [0.2,0.8]$ for each simulated AGN. In summary, the hot dust emission and its variability is
fully characterized by $a$ and $w_\mathrm{eff}$.

\begin{figure}
\begin{center}
\epsscale{1.2}
\plotone{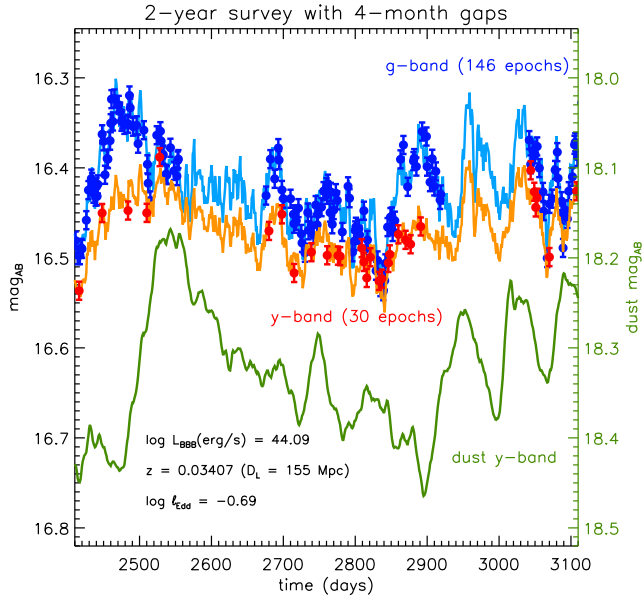}
\caption{Simulated AGN light curves in the $g$- (light blue) and
  $y$-band (light red) filters for a period of two years. The blue and
  red circles with error bars represent mock observations of these
  $g$- and $y$-band light curves, respectively, based on the
  \textit{LSST} photometric precision and an average sampling of 4
  days ($g$-band) and 17 days ($y$-band). \textbf{The green line represents
  the dust-only model light curve in the $y$-band (magnitude scale on
  right axis).} The AGN characteristics are
  listed in the top-left corner.}\label{fig:mock_obs}
\end{center}
\end{figure}

Magnitudes at
all \textit{LSST} wavebands are extracted for the combined BBB +
hot-dust emission for AGN with luminosities $L$ at distances
$D_L$. The ``mock observations'' take into account the expected
statistical and systematic errors of the \textit{LSST}\,\footnote{based on
the descriptions at
\url{http://ssg.astro.washington.edu/elsst/magsfilters.shtml} and
\url{http://ssg.astro.washington.edu/elsst/opsim.shtml?skybrightness}}. 
It is assumed that each AGN is observed once every 7 days in $u$, 3
days in $g$, 5 days in $r$, 10 days in $i$, 20 days in $z$, and 15
days in $y$.

For illustration of the proposed method, AGN properties
observed in the local universe ($z\la 0.1$) were approximated as follows:
First, a redshift is randomly picked from a $(1+z)^3$-distribution. Then, luminosities are
drawn randomly from the interval $\log L(\mathrm{erg/s}) \epsilon
[42.7,44.3]$ and adjusted by
$10\cdot z$ (quasars become more abundant with $z$). $M_\mathrm{BH}$ is determined based on
$L$ and an Eddington ratio picked randomly around the $L$-dependent mean $\log
\left<\ell_\mathrm{Edd}\right> = -1.0 + 0.3\times \log
L/\left<L\right>$ (Gaussian with standard deviation
$\sigma_{\log \ell} = 0.22\,$dex), producing an
$L$-$\ell_\mathrm{Edd}$ correlation.

\subsection{How to recover dust time lags}\label{sec:howto}

A catalog with 301 AGN has been
simulated\footnote{The catalog and analysis are available at
  http://dorm.sungrazer.org}. Example light curves in the $g$ and $y$
bands are presented in Fig.~\ref{fig:mock_obs}. The circles with error bars are
the observed epochs that will be used as input for the
reverberation experiment. In the following, a very simple
  cross-correlation approach will be used to
  successfully recover dust lags. The
intention is to provide a proof-of-concept, while optimization or tests
of better approaches \citep[e.g.][]{Che12,Che13,Zu13} are encouraged for
future studies.

\begin{figure}
\begin{center}
\epsscale{1.2}
\plotone{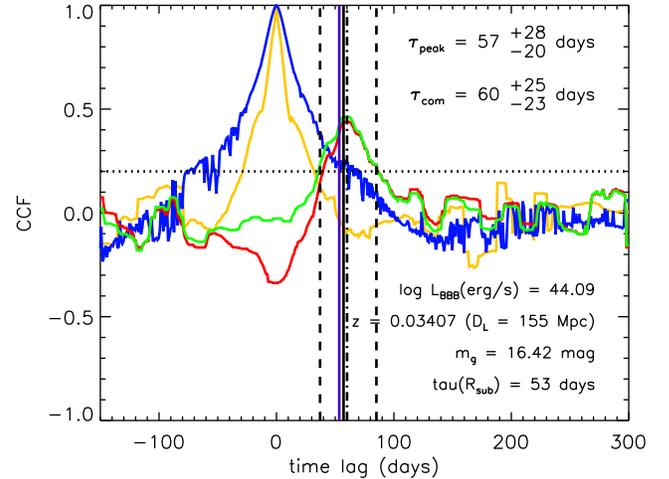}
\caption{Six-days-smoothened cross-correlation function (CCF) of an AGN
  (catalog ID 160). The red line shows the CCF between the $y-$BBB light curve and
the BBB light curve, while the blue and orange lines are
auto-correlation functions of the BBB and $y-$BBB light curves,
respectively. The green line is the $y-BBB$ CCF after subtracting
a scaled combination of the two ACFs. The dotted line marks the CCF
cutoff for automated time-delay detection. The black-solid and
-dashed-dotted lines denote the auto-detected peak time lag
$\tau_\mathrm{peak}$ and center-of-mass time lag $\tau_\mathrm{com}$,
respectively. The dashed lines mark the error regions where the CCF is
half of its peak value. For reference, the purple-solid line marks the time
lag of the sublimation radius in the input model. The AGN input properties are
listed as well.}\label{fig:ccf}
\end{center}
\end{figure}
  
First, the observed
photometric light curves in each band and with very different time
resolution and inhomogeneous coverage were resampled to a common
$\Delta t =1$\,day using the stochastic interpolation technique
described in \citep{Pet98} and \citet{Sug06}. For each band, 10 random realizations of
the resampled light curves were simulated. From the resampled $ugri$ light curves, a reference BBB light curve
was extracted. For that, the mean fluxes and standard deviations over the
10 random realizations of each band and epoch were calculated
and a simple power law $f_\mathrm{\nu} \propto \nu^\beta$ was fit to
the resulting $ugri$ fluxes at each resampled epoch. Based on this fit, a $V$-band flux at 0.55\,$\micron$
was determined. This method uses the maximum information of all bands
simultaneously and the resulting BBB light curve is very close to
the input AGN variability pattern.

In the next step, the BBB light curve was subtracted from the $y$
band fluxes. This procedure can produce negative fluxes and leaves
some BBB variability in the result because of the overestimation of
the BBB underlying the $y$ band and the wavelength-dependence of the
variability. However, as we are interested only in the time delay
signal, this does not need to be of concern. The most important result from
this procedure is that a large part of the BBB signal has been
removed in this $y-$BBB light curve.

\begin{figure*}
\begin{center}
\epsscale{0.65}
\plotone{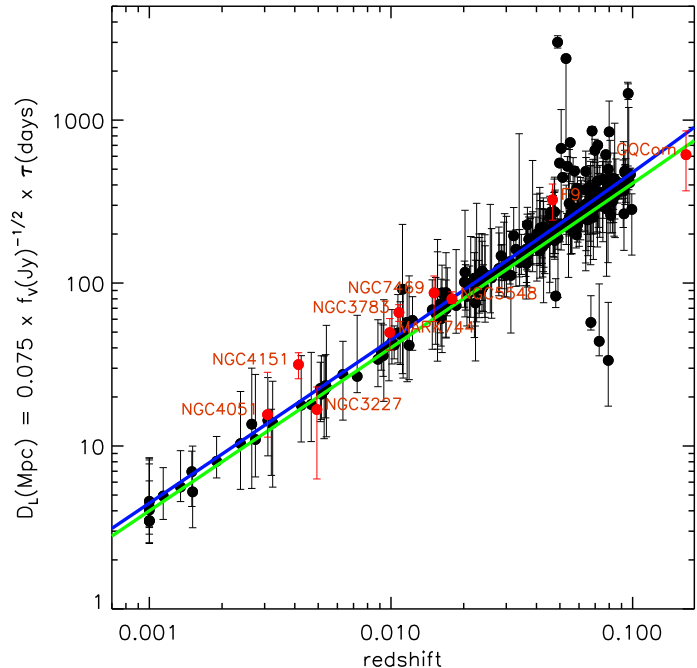}
\caption{Redshift-independent luminosity distances based on dust time
lags and AGN $V$-band flux plotted against redshift. The black circles
with error bars are AGN from the mock survey. The red points are observed AGN with known dust time
lags. The blue line shows the $z-D_\mathrm{L}$ relation for the
Planck cosmological parameters, while the green line assumes
$H_0=75$\,(km/s)/Mpc, $\Omega_m
=1$ and $\Omega_\Lambda = 0$.}\label{fig:cosmo}
\end{center}
\end{figure*}

Finally, the \textit{observed} epochs of the $y-$BBB light curve are (discretely)
cross-correlated with the BBB light curve by interpolating the BBB
flux at the observed $y$ band epochs and accounting for different
lags. An example cross-correlation function (CCF) for a ten-year
\textit{LSST} survey, smoothed with a box-car kernel with a width of 6 days, is shown in
Fig.~\ref{fig:ccf}. The CCF shows a distinct negative/anti-correlation
peak at zero lag. This peak originates from the subtraction method
discussed above. As such, it closely follows the auto-correlation
functions of the BBB and $y-$BBB light
curves. After linearily-combining both ACFs and scaling to the 0-lag
negative peak, the BBB effect on the CCF can be effectively removed.

To recover the dust lag, the highest peak in the CCF after ACF
subtraction was automatically
identified. A positive detection is considered if the peak CCF
$\ge$0.2. An error region is defined as the range over which the CCF
is at least half the peak CCF. The maximum CCF defines the time lag of the
peak $\tau_\mathrm{peak}$. A center-of-mass time lag $\tau_\mathrm{com}$ is also
determined at half the integrated CCF within the error region.

As a final remark, a direct cross-correlation between
the observed $y$ band and any optical band did not recover a
time delay, although a corresponding peak is seen when
cross-correlating the input model light curves.
In addition, the ``shifted reference'' method by
\citet{Che12} developed to recover time lags of the
BLR from photometric filters was not successful. This originates from
the fact that the target signal does not vary with the same amplitude
as the reference band and is significantly smeared, and may require the
refinement presented in \citet{Che13}.

\section{Discussion}\label{sec:discu}

\subsection{Time delays in a real survey}\label{sec:real}

The quality of lag recovery critically depends on the cadence
as well as the continuity of observations. Typically, objects will not
be observable year-round. In order to simulate this effect, annual
gaps of 2, 4, and 6 months were introduced for 1/6, 1/3, and 1/2 of
the catalog, respectively. Due to strong noise features for
lags longer than $\sim$350\,days, the CCF was analyzed only for lags
$<$300\,days. In general, lags
are detected for about 70\% or more of all AGN, independent of gap length. Out of these lags,
about 80\% are consistent within error bars with the sublimation
radius of the input model. This is arguably a high rate given the
simplicity of the lag recovery scheme and the completely blind
analysis without any human interaction. The success rates might 
be boosted with more sophisticated methods and a way to deal with the
longer lags. Note that these numbers are strictly valid only for the
specific AGN sample parameters (see Sect.~\ref{sec:simobs}).

There are some issues that merit further attention. The
  current simulations do not take into account contributions of broad
  lines. This should be a minor issue for the recovery of the BBB
  light curve given the multi-filter fitting scheme. However, broad
  components of Paschen lines in the $y$-band may impose a secondary
  lag signal onto the dust light curve \citep[see][for Balmer
  lines]{Che12}. While their contribution is probably smaller
  than the dust contribution in this band \citep[see spectra
  in][]{Lan11}, they should be part of a more advanced recovery scheme (see
  Sect~\ref{sec:howto}). Furthermore, potential
  scattered light from the BBB \citep[e.g.][]{Kis08,Gas12} and the
  Paschen continuum are also neglected, but their contribution to the total $y$-band flux
  is estimated to be of the order of 1\%.

The host flux has been neglected in the modeling (see Sect.~\ref{sec:idea}). Based on
the optical AGN/host decomposition of \citet{Ben09}, the contribution of the host to the total
flux can reach 50\% at $L_\mathrm{AGN} \la 10^{43}$\,erg/s and will
drop to $\sim$10\% for $L_\mathrm{AGN} \la 10^{45}$\,erg/s in the $V$-band. Therefore, if the host contribution was
considered, the presented simulations would apply to a sample that is
brighter by $\sim0.1-0.7$\, mag.

\subsection{Cosmological applications of dust time lags and dust emission sizes}\label{sec:cosmo}

Radius-luminosity relations open up the possibility to use AGN as ``standard candles'' in
  cosmology. Such applications have been discussed for both the BLR
  \citep{Haa11,Wat11} and the dust  \citep[e.g.][]{Kob98,Okn99b,Okn02,Yos04}. While the current AGN
  reverberation sample is larger for the BLR than for the hot dust,
  the proposed use of optical surveys may change this
  picture significantly. Fig.~\ref{fig:cosmo}
shows the 231 objects from the 301-object mock AGN catalog for which
time lags $\tau_\mathrm{peak}$ were recovered. For each of these AGN,
the $V$-band flux $f_V$ was measured and a luminosity distance $D_L$ independent of redshift was calculated as
\begin{equation}\label{eq:cosmo}
D_L(\mathrm{Mpc}) = 0.075 \times f_V(\mathrm{Jy})^{-1/2} \times
\tau_\mathrm{peak}(\mathrm{days}) \ .
\end{equation}
The scaling factor of 0.075 was obtained
by fitting $f_V^{-1/2} \cdot \tau$ to the known distances of the
objects in the mock survey (a real survey will require normalizing to the cosmic distance
ladder). The errors of individual data points are dominated by the
uncertainties in $\tau$ as long as $f_V$ can be determined with
$\la30\%$ accuracy. Overplotted are $z-D_L$-relations for a standard cosmology
according to the latest Planck results ($H_0 = 67.3$\,(km/s)/Mpc, $\Omega_m = 0.315$,
$\Omega_\Lambda = 0.685$) as well as for a universe without
cosmological constant ($H_0 = 75$\,(km/s)/Mpc, $\Omega_m = 1$,
$\Omega_\Lambda = 0$). For comparison, $K$-band reverberation-mapped
AGN from literature are also shown (NGC~3227, NGC~4051,
NGC~5548, NGC~7469: \citet{Sug06}; NGC~4151: \citet{Kos09}; Fairall 9:
\citet{Cla89}; GQ Com: \citet{Sit93}; NGC~3783: \citet{Gla92};
Mark~744: \citet{Nel96}). The limitation to
optical bands does not allow for distinguishing between different
cosmological parameters. However, such a nearby sample ($z \la 0.1$)
can be used to determine the scaling factor. Under the same
configuration as taken for the optical bands, a survey using near-IR
filters will reach $z\sim0.3$ in the $J$-band, $z\sim0.7$ in the
$H$-band, and $z\sim1.3$ in the $K$-band.

One of the disadvantages of standard candles are their
  reliance on the cosmic distance ladder. This dependence can be
  overcome with ``standard rulers'' for which an angular diameter
  distance $D_A$ is determined by comparing physical with angular
  sizes. \citet{Elv02} proposed the use of AGN BLR time lags in
  combination with spatially- and spectrally-resolved interferometry of
broad emission lines for this purpose. However, optical/near-IR interferometry of AGN needs
8-m class telescopes (e.g. at the VLTI or Keck) and is limited to
about 130m baseline lengths. A broad emission line in the
near-IR has only been successfully observed and resolved for one AGN
to-date \citep[3C273;][]{Pet12}.

Here, it is proposed that the dust
time lags may also be used as a ``standard ruler'' in combination
with directly measured near-IR angular sizes, $\rho_K$, from
interferometry to determine angular size distances $D_A$, via the
relation 
\begin{equation}
D_A(\mathrm{Mpc}) = 0.126 \times
\frac{\tau(\mathrm{days})}{\xi_\mathrm{int} \cdot
  \rho_K(\mathrm{mas})} \ .
\end{equation}
$\xi_\mathrm{int}$ corrects the
different sensitivities of the CCF and interferometry measurements to
extended dust distributions \citep[e.g.][]{Kis11a,Kis14} and may be
determined from spectral or interferometric modeling. 

For both the standard candle and ruler, time lags for the
  hot dust have to be determined. As compared to the BLR, the dust
lags are about a factor of 4 longer (comparing the $\mathrm{H\beta}$
relation in \citet{Ben09} with the dust relation in \citet{Kis07}),
resulting in longer monitoring campaigns. On the other hand, dust monitoring does not
 require spectroscopy or the use of specific narrow-band filters, but
can be executed with any telescope that has broad-band filters and
sufficient sensitivity. Moreover, the hot-dust emission and
sublimation radius are extremely uniform across AGN, e.g. showing a narrow range of color temperatures
\citep[e.g.][]{Gla04}, a universal turn-over at
$\sim1-1.2\,\micron$ \citep[e.g.][]{Neu79,Elv94}, and an emissivity close to order
unity \citep[e.g.][]{Kis11b}. This points toward simple physics (radiative
reprocessing within $\sim\micron$-sized dust grains) involved in dust
emission at $\rsub$, which arguably removes some physical
uncertainties that are inherent to other methods.

For hot dust radii as standard rulers, the emission region has
  to be spatially resolved without the need for spectral
  resolution. This has been achieved for 13 nearby AGN to-date
  \citep[$z\le0.16$; e.g.][]{Kis11a,Wei12,Kis14} with the limiting
  factor being sensitivity of IR interferometers. However,
  using a potential km-sized heterodyne array
  (e.g. the Planet Formation
  Imager\footnote{http://planetformationimager.org} with sufficient
  wavelength coverage into the $L$ and $M$ bands, space-based or on
  the ground, may allow for direct distance estimates directly into the
Hubble flow out to $z\ga1$, entirely bypassing the
cosmological distance ladder (H\"onig et al., in prep.).

\acknowledgments

\begin{footnotesize}
\textbf{Acknowledgements ---} I want to thank Darach Watson and Aaron
Barth for helpful discussions, and the anonymous referee for many
helpful comments and suggestions that improved the manuscript. The Dark Cosmology Centre is funded by
The Danish National Research Foundation.
\end{footnotesize}

\end{document}